\documentclass[aps,preprint]{revtex4}%
\usepackage{amsfonts}
\usepackage{amsmath}
\usepackage{amssymb}
\usepackage{graphicx}%
\providecommand{\U}[1]{\protect\rule{.1in}{.1in}}
\newtheorem{theorem}{Theorem}
\newtheorem{acknowledgement}[theorem]{Acknowledgement}

\begin{document}
\title{Is the proton radius puzzle an evidence of extra dimensions?}
\author{F. Dahia}
\affiliation{{\footnotesize Department of Physics, Universidade Federal da Para\'{\i}ba,
Jo\~{a}o Pessoa - PB, Brazil}}
\author{A. S. Lemos}
\affiliation{{\footnotesize Department of Physics, Universidade Federal da Para\'{\i}ba,
Jo\~{a}o Pessoa - PB, Brazil}}

\begin{abstract}
The proton charge radius inferred from muonic hydrogen spectroscopy is not
compatible with the previous value given by CODATA-2010, which, on its turn,
essentially relies on measurements of the electron-proton interaction. The
proton's new size was extracted from the 2S-2P Lamb shift in the muonic
hydrogen, which showed an energy excess of 0.3 meV in comparison to the
theoretical prediction, evaluated with the CODATA radius. Higher-dimensional
gravity is a candidate to explain this discrepancy, since the muon-proton
gravitational interaction is stronger than the electron-proton interaction
and, in the context of braneworld models, the gravitational potential can be
hugely amplified in short distances when compared to the Newtonian potential.
Motivated by these ideas, we study a muonic hydrogen confined in a thick
brane. We show that the muon-proton gravitational interaction modified by
extra dimensions can provide the additional separation of 0.3 meV between the
2S and 2P states. In this scenario, the gravitational energy depends on the
higher-dimensional Planck mass and indirectly on the brane thickness. Studying
the behavior of the gravitational energy with respect to the brane thickness
in a realistic range, we find constraints for the fundamental Planck mass that
solve the proton radius puzzle and are consistent with previous experimental bounds.

\end{abstract}
\maketitle

\section{Introduction}

The proton charge radius was determined with unprecedented precision by recent
measurements of the Lamb shift in muonic hydrogen \cite{nature,science}, the
atom formed by a muon and a proton $\left(  \mu p\right)  $. It happens that
the deduced radius $r_{p}=0.84184(67)%
\operatorname{fm}%
$ is 4\% smaller than CODATA-2010 value, $r_{p}^{CD}=0.8775(51)%
\operatorname{fm}%
$ \cite{codata} - which is inferred from hydrogen and deuteron spectroscopy
\cite{h0,h1,h2,h3,h4,h5,h6,h7} and from measurements of differential cross
section in elastic electron-proton scattering \cite{ep,ep0,ep1}. This
discrepancy of 7 standard deviations is known as the proton radius puzzle.

The proton charge radius is defined as $<r_{p}^{2}>=\int r^{2}\rho_{E}\left(
\mathbf{r}\right)  d^{3}\mathbf{r}$, where $\rho_{E}$ is the normalized
electric charge density of the proton. Based on the standard theory of
bound-state quantum electrodynamics (QED), the effects of the proton internal
structure on atomic energy spectrum can be predicted. For instance, in the
muonic hydrogen, it is expected that the contribution for the $2S_{1/2}%
-2P_{1/2}$ Lamb shift is given by \cite{science,annals}:%
\begin{equation}
\Delta E_{L}^{th}=\left[  206.0668(25)-5.2275(10)\frac{r_{p}^{2}}{%
\operatorname{fm}%
^{2}}\right]
\operatorname{meV}
\label{Lamb}%
\end{equation}
where $r_{p}$ must be expressed in femtometer. According to this formula, the
energy shift is $\Delta E_{L}^{th}\left(  r_{p}^{CD}\right)  =$ $202.0416(469)%
\operatorname{meV}%
$, when it is calculated with the CODATA-2010 radius. On the other hand, the
experimental value is extracted from the measurement of the ($2P_{3/2}%
^{F=1}-2S_{1/2}^{F=0}$) and ($2P_{3/2}^{F=2}-2S_{1/2}^{F=1}$) transitions
frequencies, $\nu_{s}$ and $\nu_{t}$ respectively, and from the formula
\cite{science,annals}:%
\begin{equation}
\Delta E_{L}^{\exp}=\frac{1}{4}h\nu_{s}+\frac{3}{4}h\nu_{t}-8.8123\left(
2\right)
\operatorname{meV}%
,
\end{equation}
where the numeric term comes from the explicit calculation of the 2P fine and
hyperfine splitting. By using the measured frequencies, $\nu_{s}%
=54611.16\left(  1.05\right)
\operatorname{GHz}%
$ \cite{science,annals} and $\nu_{t}=49881.35(65)%
\operatorname{GHz}%
$ \cite{nature,science,annals}, we find $\Delta E_{L}^{\exp}=202.3706(23)%
\operatorname{meV}%
$. The difference of $0.3290(469)%
\operatorname{meV}%
$, between the measured Lamb-shift and the predicted value, has no explanation
within the standard framework of physics. Thus the puzzle may be an indication
of a missing term in equation (\ref{Lamb}), associated with an unknown
proton-muon interaction that differs from the electron-proton interaction. New
interactions beyond the standard model have been proposed to explain the
energy excess
\cite{new,new0,new1,new2,new3,new4,new5,new6,new7,new8,new9,new10,new11,new12}%
, but there is no final conclusion yet.

Here we want to discuss an alternative explanation. As the muon is around 207
times heavier than the electron, it is reasonable to conjecture that gravity
is the missing piece in this puzzle. The problem is that the Newtonian
potential is negligible in atomic system. However, in the context of the
braneworld with large extra dimension, the gravitational potential can be much
greater in short distances. This fact has motivated us to address this issue
in the context of the braneworld models.

In the braneworld scenario, our visible Universe is a submanifold with three
spatial dimensions (the 3-brane) embedded in an ambient space of higher
dimensions (the bulk) \cite{ADD:1,ADD:2,RS:1,RS:2}. Matter and standard model
fields are confined to the brane while gravity can propagate in every
direction of the bulk. Although gravity has access to whole ambient space, the
existence of a bound zero-mode (due to a compact topology or to an appropriate
curvature of the bulk), guarantees that the Newtonian behavior is recovered
for distances greater than a characteristic length scale $\ell$ of the extra
space, making the model phenomenologically viable. In the case of compact
topology, $\ell$ is the size of the supplementary space, while in the case of
non-compact topology, $\ell$ is related to the curvature radius of the ambient
space. It follows from this picture that gravity may feel directly the effects
of extra dimensions in a length scale that could be much greater than the
scale in which matter and other fields experience the influence of extra dimensions.

Tests of the inverse square law in laboratory establish that the radius of the
extra dimension should be smaller than $44%
\operatorname{\mu m}%
$ \cite{LONG,Hoyle:2001,Hoyle:2004,Hoyle:2007,review}. This is the tightest
constraint for models with only one extra dimension. When the number of extra
dimensions is greater, the most stringent constraints come from astrophysics
\cite{SN,neutronstar} and high energy particle collisions
\cite{colliders,lhc,monojet,landsberg}.

If the gravitational field obeys the Gauss law in the bulk, in the weak field
limit, then the gravitational potential of a point-like mass behaves as
$\left(  G_{n}m\right)  /r^{n+1}$ for $r<<\ell$, where $G_{n}$ is the
gravitational constant defined in a space with $n$ extra dimensions. The
relation between the Newtonian constant, $G$, and $G_{n}$ is given by
$G_{n}\sim G\ell^{n}$, in magnitude order. Therefore, in comparison with the
Newtonian potential, the extra-dimensional version is amplified by a factor of
the order of $\left(  \ell/r\right)  ^{n}$ in short distances. This property
has motivated the study of the gravitational interaction in atomic and
molecular systems as a way of obtaining empirical bounds for the number and
size of extra dimensions
\cite{specforxdim1,atomicspec,Li,Wang,specforxdim2,molecule,dahia}.
Considering that the gravitational interaction is a small term of the atomic
Hamiltonian, we find that the gravitational energy is proportional to the mean
value of $r^{-\left(  n+1\right)  }$ in the atomic state. However, this
average diverges for $S-$states, when $n>2$. To avoid this problem, some
authors introduce a cut-off radius to perform the calculations
\cite{specforxdim1,atomicspec,Li,Wang,specforxdim2}. As a consequence, the
results become dependent on an arbitrary parameter. Previous attempts of
solving the proton radius puzzle by means of the extra-dimensional gravity
also resorted to a cut-off radius \cite{Li,Wang}.

In thick brane scenario the divergence problem is naturally solved. The origin
of the divergences is the fact that a delta-like confinement in the brane is a
singular distribution from the viewpoint of the bulk \cite{dahia,effbrane}.
However, in a thick brane scenario, the confined particles are described by a
regular wave function with a non-null width in the transversal directions.
This width should be less than the brane thickness and its value is related to
the strength of the confinement. As the width is non-null, the divergence
problem naturally disappears.

Considering the muonic hydrogen in this scenario, we find the energy shift of
the atomic states caused by the muon-proton gravitational interaction. Based
on these calculations, we show that the gravitational energy can account for
the energy excess of the measured Lamb shift, solving, in this way, the proton
radius puzzle. This condition determines some constraints for the
higher-dimensional Planck mass which are consistent with previous empirical bounds.

\section{The gravitational energy of an atom in a thick brane}

In the field-theory framework, the brane can be seen as a topological defect
capable of trapping matter inside its core \cite{rubakov}. As an illustration,
we can mention a domain wall in (4+1)-dimensions that separates two vacuum
states of a scalar field $\phi$ along the extra dimension $z$ \cite{rubakov}.
In this configuration, the scalar field can confine matter in the center of
the wall by means of a Yukawa-type interaction with Dirac spinors. Under the
influence of this interaction, the zero-mode state is described by the
following wave function:%
\begin{equation}
\Psi\left(  \mathbf{x},z\right)  =\exp\left[  -\beta\int_{0}^{z}\phi
_{0}\left(  y\right)  dy\right]  \psi\left(  \mathbf{x}\right)  , \label{wave}%
\end{equation}
where $\beta$ is the coupling constant, $\psi\left(  \mathbf{x}\right)  $
represents a free spinor in the $\left(  3+1\right)  $-dimensions, $\phi_{0}=$
$\eta\tanh\left(  z/\varepsilon\right)  $ is the scalar in a domain wall
configuration interpolating between two vacua $\pm\eta$ of the scalar field.
This wave function has a peak at the center of the brane $\left(  z=0\right)
$ and decreases exponentially in the transverse direction. The parameter
$\varepsilon$ can be seen as a measure of the brane thickness, which must be
smaller than $10^{-19}%
\operatorname{m}%
$ to be consistent with current experimental constraints \cite{ADD:1,lhc}.

Confinement mechanism for matter in topological defects of greater codimension
can be also formulated in a similar away. Based on the previous example, it is
reasonable to expect that the wave function of localized particles can be
written as $\Psi\left(  \mathbf{r},\mathbf{z}\right)  =\chi\left(
\mathbf{z}\right)  \psi\left(  \mathbf{r}\right)  $, where $\chi\left(
\mathbf{z}\right)  $ is some normalized function defined in the supplementary
space of $n$-dimensions, concentrated around the origin.

In this context, let us now study the gravitational potential produced by a
confined particle in the thick brane. As we are assuming that $\ell
>>\varepsilon$, then we have to consider the direct effects of the extra
dimensions on the gravitational potential. To take this into account, we will
admit that the static gravitational field satisfies the Gauss law in the bulk.
Thus, in the case of a flat supplementary space with a compact topology, the
exact potential of a point-like mass $M$ lying in the origin of the coordinate
system and evaluated at the position $\mathbf{R}=\left(  \mathbf{r}%
,\mathbf{z}\right)  $ can be written as \cite{kehagias}:
\begin{equation}
V\left(  \mathbf{R}\right)  =-\frac{G_{n}M}{R^{n+1}}-\sum_{i}\frac{G_{n}%
M}{\left\vert \mathbf{R}-\mathbf{R}_{i}^{\prime}\right\vert ^{n+1}},
\label{potential}%
\end{equation}
where the sum spans the topological images of $M$ in the covering space of the
extra-dimensional manifold and $R=\left\vert \mathbf{R}\right\vert $. The
exact position $\mathbf{R}_{i}^{\prime}$ of the mirror images depends on the
topology of the supplementary space. For instance, in the case of a flat
$n-$torus with size $\ell$, we have $\mathbf{R}_{i}=\ell\left(
0,0,0,\mathbf{k}_{i}\right)  $, where $\mathbf{k}_{i}$ is a vector with $n$
integer number as components. The gravitational potential (\ref{potential})
reduces to the Newtonian potential $-GM/r$ in the far zone $\left(
r>>\ell\right)  $ \cite{kehagias}.

Regarding the influence of the gravitational potential on the energy spectrum
of the muonic hydrogen, the topological images can be neglected, since the
contribution they give is lesser than the empirical error of the $\mu p$
experiment (see appendix). Therefore, to calculate the proton gravitational
potential, $\phi$, we may use the approximate Green function $-GM/R^{n+1}$,
which is weaker than the real potential of a point-like mass. So, assuming
that the proton mass $m_{p}$ is distributed on the spatial extension of the
nucleus, the proton gravitational potential is
\begin{equation}
\phi\left(  \mathbf{R}\right)  =-G_{n}\int\frac{\rho_{M}\left(  \mathbf{R}%
^{\prime}\right)  }{\left\vert \mathbf{R}-\mathbf{R}^{\prime}\right\vert
^{n+1}}d^{3+n}\mathbf{R}^{\prime}, \label{protonpotential}%
\end{equation}
where the mass density is $\rho_{M}=\left\vert \Psi_{p}\right\vert ^{2}m_{p}$
and $\Psi_{p}\left(  \mathbf{r},\mathbf{z}\right)  =\chi_{p}\left(
\mathbf{z}\right)  \psi_{p}\left(  \mathbf{r}\right)  $ is the
higher-dimensional wave function of the proton.

The muon-proton gravitational interaction, which is described by the
Hamiltonian $H_{G}=m_{\mu}\phi$ (where $m_{\mu}$ is the muon mass), modifies
the muonic hydrogen spectrum. Assuming that $H_{G}$ is a small term of the
atomic Hamiltonian, the energy shift can be calculated by the perturbation
method for each state. In the first order, the energy correction is
$\,\left\langle m_{\mu}\phi\right\rangle _{\Psi}$, i.e., the mean value of the
gravitational energy in the state $\Psi$. By using Eq. (\ref{protonpotential}%
), we can write the energy shift as:
\begin{equation}
\delta E_{\psi}^{g}=-G_{n}m_{p}m_{\mu}\int\frac{\left\vert \Psi_{p}\right\vert
^{2}\left\vert \Psi_{\mu}\right\vert ^{2}}{\left\vert \mathbf{R}%
-\mathbf{R}^{\prime}\right\vert ^{n+1}}d^{3+n}\mathbf{R}d^{3+n}\mathbf{R}%
^{\prime}, \label{energy}%
\end{equation}
where the higher-dimensional wave function of the muon (more precisely, the
reduced particle) $\Psi_{\mu}\left(  \mathbf{r},\mathbf{z}\right)  $ is the
product of the extra-dimensional part $\chi_{\mu}\left(  \mathbf{z}\right)  $
and the solutions $\psi_{\mu}\left(  \mathbf{r}\right)  $ of the
Schr\"{o}dinger equation for the muonic hydrogen.

To calculate (\ref{energy}), we shall assume that the proton mass is uniformly
distributed inside the nucleus. This means that the 3-dimensional part,
$\psi_{p}\left(  \mathbf{r}\right)  $, is constant in the spatial extension of
the nucleus and zero outside $\left(  r>r_{p}^{CD}\right)  $. In equation
(\ref{energy}), the major contribution comes from the integral in the interior
region of the nucleus. For $S-$states, equation (\ref{energy}) yields%
\begin{equation}
\delta E_{S}^{g}=-\gamma_{n}\frac{G_{n}m_{p}m_{\mu}}{\sigma^{n-2}}\left\vert
\psi_{S}\left(  0\right)  \right\vert ^{2}\left[  1-\frac{3}{2}\frac{r_{p}%
}{a_{0}}+O\left(  r_{p}^{2}/a_{0}^{2}\right)  \right]  \left[  1+O\left(
\sigma/r_{p}\right)  \right]  , \label{shift}%
\end{equation}
where $a_{0}$ is the Bohr radius of the muonic hydrogen, $\psi_{S}\left(
0\right)  $ is the wave function of a $S$-state evaluated in the origin and
$\gamma_{n}$ is a numeric factor whose value depends on the number of extra
dimension. For instance, $\gamma_{3}=2\pi^{3/2},\gamma_{4}=4\pi/3,\gamma
_{5}=\pi^{3/2}/3$ and $\gamma_{6}=$ $4\pi/15$. The gravitational energy
depends on how tight is the confinement in the thick brane. In fact, $\sigma$
is associated to the spatial distribution of the particles in the transverse
direction. This parameter is defined as:%
\begin{equation}
\frac{1}{\sigma^{m}}\equiv\frac{\Gamma\left(  n/2\right)  }{\Gamma\left(
\frac{n-m}{2}\right)  }\int\frac{\left\vert \chi_{p}\left(  \mathbf{z}%
_{1}\right)  \right\vert ^{2}\left\vert \chi_{\mu}\left(  \mathbf{z}%
_{2}\right)  \right\vert ^{2}}{\left\vert \mathbf{z}_{1}-\mathbf{z}%
_{2}\right\vert ^{m}}d^{n}\mathbf{z}_{1}d^{n}\mathbf{z}_{2}, \label{sigma}%
\end{equation}
where $m$ is a positive integer that should satisfy the condition
$m\leq\left(  n-1\right)  $ and $\Gamma$ stands for gamma function. If $\chi$
is a Gaussian function, then $\sigma$ coincides with the standard deviation of
the Gaussian distribution. For the sake of consistency, $\sigma$ should be
smaller than the brane thickness.

Equation (\ref{shift}) is valid for $n>2$. Here we do not discuss the cases
$n=1$ and $n=2$, once the atomic gravitational energy are not strong enough to
explain the proton radius puzzle in those dimensions. The integral of equation
(\ref{energy}) in the external region is smaller than (\ref{shift}) by a
factor of the order of $\sigma/r_{p}$, which is lesser than $10^{-5}$ for
realistic branes with $\varepsilon\leq10^{-20}%
\operatorname{m}%
$. On its turn, for $P$-states, the gravitational contribution is smaller than
(\ref{shift}) by a factor of the order of $r_{p}^{2}/a_{0}^{2}$.

\section{The additional energy in the Lamb shift}

As we have already mentioned, in comparison with the predicted Lamb shift
$\Delta E_{L}^{th}\left(  r_{p}^{CD}\right)  $, the measured value $\Delta
E_{L}^{\exp}$ has an excess of $0.3290(469)%
\operatorname{meV}%
$. The higher-dimensional gravity can explain this excess in a consistent way.
Due to the proton-muon gravitational interaction, the energy of $2S$-level
decreases by the amount $\delta E_{2S}^{g}=-\gamma_{n}G_{n}m_{p}m_{\mu}\left(
1-3r_{p}/2a_{0}\right)  /(8\pi a_{0}^{3}\sigma^{n-2})$, according to equation
(\ref{shift}). On the other hand, the effect on $2P-$level is smaller by a
factor of the order of $10^{-5}$, therefore, it is negligible within the
precision of $10^{-7}%
\operatorname{eV}%
$ of the muonic hydrogen experiment \cite{annals}. Thus, the gravitational
interaction is responsible for an additional enlargement between the levels
$2P-2S$ given by $\left\vert \delta E_{2S}^{g}\right\vert $. The puzzle would
be solved if $\left\vert \delta E_{2S}^{g}\right\vert =0.3290(469)%
\operatorname{meV}%
$. This condition implies a relation between $G_{n}$ and $\sigma$, which, in
terms of the fundamental Planck mass $M_{D}$ of the higher-dimensional space,
as defined in ref. \cite{colliders}, can be written as:%
\begin{equation}
\left[  \frac{\left(  \hbar/c\right)  ^{n}\hbar c}{(n+2)}\frac{\Gamma\left(
\frac{n+3}{2}\right)  }{2\pi^{\left(  n+3\right)  /2}}\frac{\left(
2\pi\right)  ^{n}}{M_{D}^{n+2}}\right]  \frac{\gamma_{n}}{8\pi}\frac
{m_{p}m_{\mu}}{a_{0}^{3}\sigma^{n-2}}\left(  1-\frac{3r_{p}}{2a_{0}}\right)
=0.3290(469)%
\operatorname{meV}%
, \label{planckmass}%
\end{equation}
where $G_{n}$ was substituted by the term in the bracket. Figure (1) shows a
numerical analysis of equation (\ref{planckmass}) for four cases $n=3,4,5$ and
$6$. The constraints yield the required values of $M_{D}$, in the range
$10^{-35}%
\operatorname{m}%
$ $\leq\sigma\leq10^{-20}%
\operatorname{m}%
$, in order to solve the proton radius puzzle. As we can see, thinner branes
$-$ which imply tighter confinements, i.e., smaller $\sigma$ $-$ demand higher
values for the fundamental Planck mass. The uncertainty on the
higher-dimensional Planck mass at one standard deviation level is $\delta
M_{D}/M_{D}=0.1426/(n+2)$ for a fixed $\sigma$, and it is too narrow to be
seen in Figure 1.%
\begin{center}
\includegraphics[
height=2.7259in,
width=2.9179in
]%
{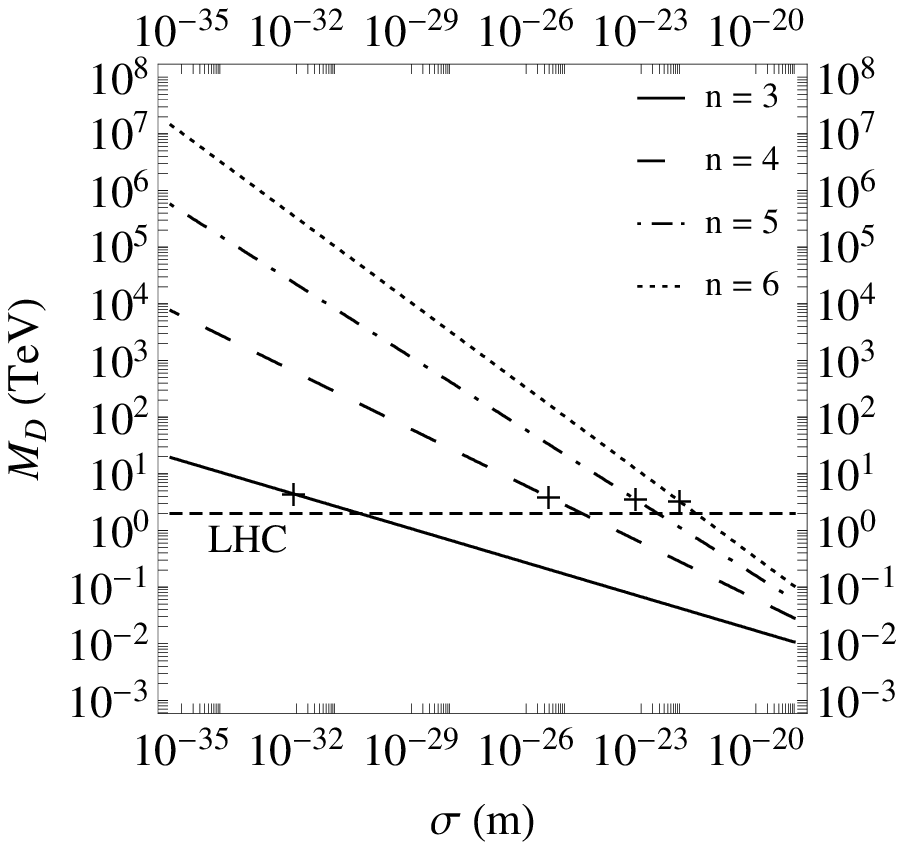}%
\\
The required values of higher-dimensional Planck mass (in natural units) to
explain the proton radius puzzle, in terms of the confinement parameter
$\sigma$. The region below the 2 TeV line is excluded from the analysis of
monophoton events in proton-proton collision at LHC. The + signs are lower
bounds from data on the monojet events at LHC.
\end{center}
Let us now compare these constraints with other experimental bounds. Direct
tests on deviation of the inverse square law at short distances, based on
modern versions of torsion-balance instrument, have been used with the purpose
of searching for signals of extra dimensions. In these experiments, the
modified gravitational potential is parameterized as $GM/r\left(  1+\alpha
e^{-r/\lambda}\right)  $, where, in the ADD model, $\alpha=8n/3$ and $\lambda$
is equal to the radius $R$ of extra dimensions
\cite{Hoyle:2001,Hoyle:2004,Hoyle:2007,review}. From the empirical constraints
on $\alpha$ and $\lambda$, upper bounds for $R$ are inferred for each value of
$n.$ For instance, for $n=1$ and $n=2$, the data imply that $R<$ $44%
\operatorname{\mu m}%
$ and $R<$ $37%
\operatorname{\mu m}%
$, respectively, which corresponds (see, the relation between $R$ and $M_{D}$
in the appendix) to $M_{D}>3.6$ TeV for $n=2$ \cite{Hoyle:2007,pdgAlex}. For
greater codimensions, the experimental limits are much below than TeV scale
and, therefore, compatible with constraints of Figure 1. If the modification
of the gravitational potential is due to radion exchange between matter,
instead of graviton exchange, the parameters have different meaning. In this
case, $\alpha=n/\left(  n+2\right)  $ and $\lambda$ is the Compton wavelength
of the radion, which is related to $M_{\ast}$ (the unification scale
\cite{radion}) by the formula $\lambda^{2}\sim\left(  \hbar^{3}/cGM_{\ast}%
^{4}\right)  $ \cite{radion,adelbergREV,antoniadis}. According to Ref.
\cite{radion}, the limits go from $M_{\ast}>5.7$ TeV $(n=1)$ to $M_{\ast}>6.4$
TeV $(n=6).$ Although the exact relation between $\lambda$ and the fundamental
Planck mass depends on the stabilization mechanism of the radion
\cite{antoniadis}, there is plenty space to accommodate these bounds in Figure
1, for $n>3$.

Astrophysical and cosmological constraints are strong for $n\leq4$ and are
derived from the implications of the supposed production of the KK-gravitons
in stars \cite{SN,neutronstar,pdgAlex,pdgGiudice}. In this context, the most
stringent bound is obtained from the analysis of this process in supernovae
explosions. In the ADD higher-dimensional model, an old remnant neutron star
is surrounded by trapped KK gravitons which slowly decay into photons. A
fraction of them is absorbed by the neutron star causing its heating. As the
excess heat is not observed, constraints can be obtained. Data from PSR
J09521+0755 demand that $M_{D}>76$ TeV for $n=3$ \cite{neutronstar,pdgAlex}.
In principle, this limit would rule out the case $n=3$ of our analysis in
Figure 1. However, it is important to have in mind that astrophysics bounds
could be evaded by some mechanism that provides an extra mass for KK gravitons
\cite{pdgGiudice,astro}.

When the number of extra dimensions is greater than four, the tightest
constraints of the fundamental Planck mass comes from high-energy collisions.
Recent analysis on monophoton events in proton-proton collision at $\sqrt
{s}=7$ TeV and $\sqrt{s}=8$ TeV in the LHC \cite{lhc} determines that
$M_{D}>2$ TeV, for $n=3,...,6$. In Figure (1), this lower bound is represented
by the horizontal line. On its turn, the analysis of monojet events in
proton-proton collision at $\sqrt{s}=8$ TeV provides stronger constraints.
Considering the LO cross-section for direct graviton emission in the
collision, the lower bounds for $M_{D}$ in TeV are: $4.38$ $\left(
n=3\right)  $, $3.86$ $\left(  n=4\right)  ,$ $3.55$ $\left(  n=5\right)  $
and $3.26$ $\left(  n=6\right)  $ \cite{monojet,landsberg}. The bounds are
indicated in Figure 1 by a $+$ sign. Above these values, constraints of Figure
1 are compatible with the collider limits too.

Finally let us now compare our results with other spectroscopy data. In a
previous work \cite{dahia}, considering a hydrogen stuck in a thick brane, we
determined lower bounds for the higher-dimensional Planck mass from the
$\left(  2S-1S\right)  $-transition. The limits from H spectroscopy are weaker
than those necessary to solve the proton radius puzzle. This means that,
considering the current constraints of $M_{D}$ as shown in Figure 1, the
gravitational energy is capable of explaining the additional difference
between the $2S_{1/2}$ and $2P_{1/2}$ states of $\mu p$, but it is still
hidden in the H spectrum. The reason is that, according to Eq. (\ref{shift}),
the atomic energy due to the gravitational interaction depends on the lepton
mass to the fourth order, $m^{4}$, approximately, since the energy is
proportional to $m/a_{0}^{3}$ and $a_{0}$ is defined in terms of atomic
reduced mass. Thus, the gravitational energy of the hydrogen is almost
$\left(  200\right)  ^{4}$ times smaller than that of $\mu p$, assuming that
the confinement of both atoms is similar, i.e., $\sigma_{H}\simeq\sigma_{\mu
p}$. Therefore, the constraints from the muonic hydrogen are also compatible
with the most precise spectroscopic data available, which is provided by the
hydrogen spectrum.

So, based on these considerations, we can conclude that there are regions in
Figure 1 in which the required values of $M_{D}$ to solve the proton radius
puzzle satisfy all previous experimental bounds.

At this point, it is important to emphasize that the calculations we have done
here are based on the classical behavior of gravity. However, as pointed out
in Ref. \cite{colliders}, quantum-gravity effects may become significant in a
length scale of the order of $l_{D}$ (the Planck length defined in the
higher-dimensional space, which is given by $l_{D}=\left(  \hbar/c\right)
M_{D}^{-1}$ ) or even in a greater scale, depending on the fundamental theory
of gravity, not yet known. If this is the case, then unpredicted phenomena
could distort or even overshadow the classical effects we have investigated here.

However, according to \cite{effgravity}, if the theory of General Relativity
is considered as an effective theory, then it is possible to estimate quantum
corrections to the gravitational potential energy. In three-dimensional space,
if $d$ is the distance between particles with mass $M$ and $m$, then the
classical potential energy is given by the Newtonian term $GMm/d$ and quantum
contributions are smaller by a factor of the order of $\left(  l_{p}/d\right)
^{2}$, where $l_{p}$ is the usual Planck length. In the higher-dimensional
case, the classical term is $G_{n}Mm/d^{n+1}$, and, according to dimensional
analysis, the quantum corrections would be of the order of $\left(
l_{D}/d\right)  ^{n+2}$. Of course, in the muonic hydrogen, proton and muon
cannot be considered as point-like particles. Nevertheless, it is instructive
to define an effective distance between them, in the extra-dimensional space,
$d_{eff}$, by writing the atomic gravitational energy as $G_{n}m_{p}m_{\mu
}/d_{eff}^{n+1}$. Now, from equation (\ref{planckmass}), $d_{eff}$ can be
estimated. Comparing it with the fundamental Planck length, we verify that the
ratio $\left(  l_{D}/d_{eff}\right)  ^{n+2}$ depends on $n$ and $\sigma$, but,
for any dimension and for any value of $\sigma$ investigated here, it is
smaller than $10^{-4}$. Thus, if $d_{eff}$ is the relevant characteristic
length scale of the system concerning its gravitational interaction, then we
can expect that the classical contribution will be the leading gravitational
influence in this system within the braneworld scenario we are considering
here. However, as the fundamental quantum gravity theory is not known, only
experiments can answer this question.

\section{Final remarks}

In the thick brane scenario, the direct influence of extra dimensions on
gravity arises in a length scale $\ell$ that may be much greater than the
scale in which standard model fields feel directly the effects of
supplementary space. It happens that the modified gravitational potential is
amplified in small distances $\left(  r<<\ell\right)  $ when compared to the
Newtonian potential. In this context, we found that the proton-muon
gravitational interaction can explain the excess of 0.3 $%
\operatorname{meV}%
$ in the Lamb shift of muonic hydrogen, provided that the fundamental Planck
mass satisfies some constraints. In Figure (1), we can find constraints for
$M_{D}$ which solve the proton radius puzzle without violating any previous
empirical bound.

In the muonic hydrogen experiment, the 2S hyperfine splitting (2S-HFS) was
investigated too \cite{annals}. In the leading order, the proton structure
affects 2S-HFS by means of the so-called Zemach radius, which is defined in
terms of the convolution between the electric and magnetic distribution of
proton. Within the current precision, the gravitational energy does not change
2S-HFS. This result is consistent with the fact that measurements of Zemach
radius extracted from the muonic hydrogen and from H spectroscopy are compatible.

The proton radius puzzle may be the first empirical evidence of the existence
of hidden dimensions. In view of this exciting implication, the model must be
tested further. It is important to investigate the theoretical predictions for
other transitions. As an example, let us mention the $2S-1S$ transition. In
the muonic hydrogen, it is expected an extra energy of $2.1$ $%
\operatorname{meV}%
$ in this transition. On its turn, in the electronic hydrogen, assuming that
$\sigma_{H}\simeq\sigma_{\mu p}$, this model predicts that the $\left(
2S-1S\right)  -$transition frequency should exhibit an excess of 420 $%
\operatorname{Hz}%
$, which is greater than the experimental error of $10$ $%
\operatorname{Hz}%
$ \cite{h0}. In spite of this, the extra-dimensional effect is still hidden in
H spectroscopy because of uncertainties related to the measurement of the
proton radius, which corresponds to $32$ $%
\operatorname{kHz}%
$ \cite{pachuki}. Thus, to reveal the traces of extra dimensions in the
$\left(  2S-1S\right)  $-transition of the hydrogen, the precision of $r^{CD}$
should be improved.

In contrast with other alternatives, a distinguishable characteristic of this
model is the universality of the effects. All atoms are affected by extra
dimensions through the modification of the gravitational interaction.
Moreover, in the case of hydrogen-like atoms, equation (\ref{shift}) predicts
a peculiar dependence of the gravitational energy on the mass of the atomic
particles. The energy shift of any $S$-state is proportional to $\left(
Mm\right)  ^{4}/\left(  m+M\right)  ^{3}$, where $M$ is the nucleus mass and
$m$ is the mass of the orbiting particle. Assuming the confinement in the
brane is similar for all atoms, we can estimate the energy shift caused by
extra dimensions in any exotic hydrogen-like atom. Experimental confirmation
of the predicted behavior could be an indication of the existence of extra dimensions.

\section{Appendix}

Let us consider equation (\ref{potential}). Here we want to demonstrate that
the potential generated by the topological images, $V_{im}$, is negligible
within the $\mu p$ experiment precision. The value of $V_{im}$ depends on the
point $\mathbf{R=}\left(  \mathbf{r},\mathbf{z}\right)  $. We obtain an upper
bound for $\left\vert V_{im}\right\vert $ by evaluating each term of the
series in an appropriate point $\mathbf{R}_{\min}$ of the ambient space of
least distance from the corresponding topological image (located in the
covering space). First, notice that for all the points $\mathbf{R}_{\min}$, we
should have $\mathbf{r}=0$. Thus:%
\begin{equation}
\left\vert V_{im}\right\vert =\sum_{i}\frac{G_{n}M}{\left\vert \mathbf{R}%
-\mathbf{R}_{i}^{\prime}\right\vert ^{n+1}}\leq\sum_{i}\frac{G_{n}%
M}{\left\vert \mathbf{z}-\ell\mathbf{k}_{i}\right\vert ^{\left(  n+1\right)
}}%
\end{equation}
If the supplementary space is a flat $n$-torus with sides of length $\ell$,
then $-\ell/2\leq z_{i}\leq$ $\ell/2$. Let us consider that the real mass $M$
is in the center of this space, which we will denote by $T_{0}\left(
\ell\right)  $. The covering space, $%
\mathbb{R}
^{n}$, can be viewed as if it were filled by cells that are copies of
$T_{0}\left(  \ell\right)  $. Now consider a mirror image $i$ of $M$ that
belongs to another cell. We want to determine the least distance from $i$ to
$T_{0}\left(  \ell\right)  $. Notice that the fundamental cell $T_{0}\left(
\ell\right)  $ is inside a ball $B$ of radius $d=\sqrt{n}\ell/2$ (the
semi-diagonal of $T_{0}\left(  \ell\right)  $). For $n<8$, only the first
neighbors, whose distance to the center of $T_{0}\left(  \ell\right)  $ is
$\ell$, are inside this ball. And, with respect to them, the least distance to
$T_{0}\left(  \ell\right)  $ is $\ell/2$.

Now let us consider images which are outside $B$. The distance from $i$, whose
position is $\ell\mathbf{k}_{i}$, to any point of $T_{0}\left(  \ell\right)  $
is greater than the radial distance from $i$ to the surface of $B$, i.e.,
$\left\vert \mathbf{z}-\ell\mathbf{k}_{i}\right\vert \geq\ell k_{i}-d$, for
any $\mathbf{z}\in T_{0}$ and $k_{i}=\left\vert \mathbf{k}_{i}\right\vert >1$.
Therefore, $\ell k_{i}-d$ is a lower estimate of the least distance from $i$
to $T_{0}\left(  \ell\right)  $. There are $2n$ images with $k_{i}=1$. The
next neighbors have $k_{i}=\sqrt{2}$. Thus, separating the contributions from
first neighbors $\left(  k_{i}=1\right)  $, we may write:%

\begin{equation}
\left\vert V_{im}\right\vert \leq\frac{2nG_{n}M}{\left(  \ell/2\right)
^{n+1}}+\frac{G_{n}M}{\ell^{n+1}}\sum_{k_{i}\geqslant\sqrt{2}}\frac{1}{\left(
k_{i}-\sqrt{n}/2\right)  ^{\left(  n+1\right)  }}.
\end{equation}
To estimate this quantity, let us define $T_{i}\left(  1\right)  $ $-$ the
symmetric $n$-torus of unity size with center at the image $i$. Each term
within the summation sign can be interpreted as the volume of a column above
$T_{i}\left(  1\right)  $ and whose height is given by the step function
$f(k_{i})\equiv\left(  k_{i}-\sqrt{n}/2\right)  ^{-\left(  n+1\right)  }$ .
Now we introduce the continuous function $g\left(  x\right)  =\left(
x-\sqrt{n}\right)  ^{-\left(  n+1\right)  }$, where $\mathbf{x}$ is the
position vector in the covering space. The semi-diagonal of $T_{i}\left(
1\right)  $ measures $\sqrt{n}/2$. Therefore, for every $\mathbf{x}\in
T_{i}\left(  1\right)  $, $x\leq\left(  k_{i}+\sqrt{n}/2\right)  $. Now, as
$g\left(  x\right)  $ is a decreasing function, then $g\left(  x\right)  \geq
g\left(  k_{i}+\sqrt{n}/2\right)  =f\left(  k_{i}\right)  $ inside the cell
$T_{i}\left(  1\right)  $. Thus, the integral of $g\left(  x\right)  $ in the
region $T_{i}\left(  1\right)  $ is an upper estimate for $f(k_{i})$. For the
sake of consistency, we should have $x>\sqrt{n}$, according to the definition
of $g\left(  x\right)  $. On the other hand, as $x\geq\left(  k_{i}-\sqrt
{n}/2\right)  $ for $x\in T_{i}\left(  1\right)  $, then we may conclude that
the previous analysis are valid for $k_{i}>3\sqrt{n}/2$, i.e., only for cells
whose center is separated from the origin by a distance greater than
$3\sqrt{n}/2$. A possible choice is $k_{i}=2\sqrt{n}$. Closer cells should be
taken separately. Thus, computing the contributions given by the first
neighbors ($k_{i}=1$), by the images at intermediary positions $\left(
\sqrt{2}\leq k_{i}\leq2\sqrt{n}\right)  $ and from the most distant images
$\left(  k_{i}>2\sqrt{n}\right)  $, we may write the potential $V_{im}$ as:
\begin{equation}
\left\vert V_{im}\right\vert \leq\frac{G_{n}M}{\ell^{n+1}}\left(
2^{n+2}n+\sum_{\sqrt{2}\leq k_{i}\leq2\sqrt{n}}\frac{1}{\left(  k_{i}-\sqrt
{n}/2\right)  ^{\left(  n+1\right)  }}+\frac{1}{n^{3/2}}\frac{2\pi^{n/2}%
}{\Gamma\left(  n/2\right)  }\left(  3^{n}-1\right)  \right)  , \label{Vimage}%
\end{equation}
where we have employed the following estimation:
\begin{equation}
\sum_{k_{i}>2\sqrt{n}}\frac{1}{\left(  k_{i}-\sqrt{n}/2\right)  ^{\left(
n+1\right)  }}\leq\int\limits_{x>3\sqrt{n}/2}\frac{d^{n}\mathbf{x}}{\left(
x-\sqrt{n}\right)  ^{\left(  n+1\right)  }}=\frac{1}{n^{3/2}}\frac{2\pi^{n/2}%
}{\Gamma\left(  n/2\right)  }\left(  3^{n}-1\right)
\end{equation}

Therefore, in the case of muonic hydrogen, the gravitational energy due to the
topological images is lesser than $\left(  G_{n}m_{p}m_{\mu}/\ell
^{n+1}\right)  F_{n}$, where $F_{n}$ is the function defined from
(\ref{Vimage}), which depends only on the number of extra dimensions and that
can be explicitly calculated for $n<$ $8$. In magnitude order, $G_{n}\sim
G\ell^{n}$, then, in terms of the Newtonian gravitational constant, we may
write $\left\vert V_{im}\right\vert \leq\left(  Gm_{p}m_{\mu}/\ell\right)
\tilde{F}_{n}$. As the precision of the muonic hydrogen experiment is $10^{-7}%
\operatorname{eV}%
$, the effect of the topological images would be detectable only if
$\ell\lesssim10^{-36}%
\operatorname{m}%
$. However, as we shall see, the order of $\ell$ is much greater according to
our constraints. To verify this, let us estimate the size of the extra
dimensions. The relation between $\ell$ and the higher-dimensional Planck mass
($M_{D}^{2+n}$) is given by $G^{-1}=8\pi\mathcal{R}^{n}\mathcal{M}_{D}^{2+n}$,
where $\mathcal{R}=\left(  \ell/2\pi\right)  $ is the radius of the extra
dimensions and $\mathcal{M}_{D}^{2+n}=M_{D}^{2+n}/[\left(  \hbar/c\right)
^{n}\hbar c]$. This relation allows us to estimate the size of extra
dimensions from the constraints of $M_{D}$ given by Figure 1. Empirical bounds
for the torus radius $\left(  \ell/2\pi\right)  $ in the cases $n=3,$..., $6$
are shown in Figure 2. By using these values, we can explicitly check that the
contribution of the topological images is negligible indeed.%
\begin{center}
\includegraphics[
height=2.6351in,
width=2.9179in
]%
{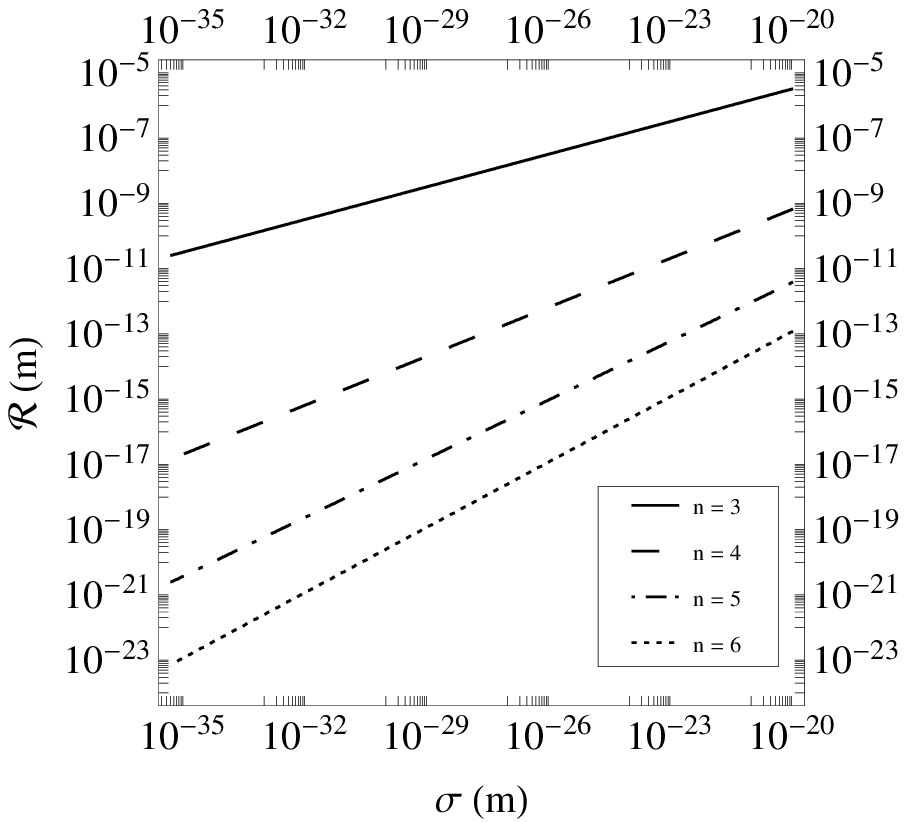}%
\\
Constraints for the radius $\left(  \mathcal{R}=\ell/2\pi\right)  $ of the
supplementary space (a flat n-torus) in terms of the confinement parameter
$\sigma$.
\end{center}

\begin{acknowledgement}
A. S. Lemos thanks CAPES for financial support.
\end{acknowledgement}

\end{document}